\documentclass{emulateapj}

\usepackage{natbib}

\usepackage{graphicx}
\usepackage[space]{grffile}
\usepackage{latexsym}
\usepackage{amsfonts,amsmath,amssymb}
\usepackage{url}
\usepackage[utf8]{inputenc}
\usepackage{fancyref}
\usepackage{hyperref}
\hypersetup{colorlinks=false,pdfborder={0 0 0},}
\usepackage{textcomp}
\usepackage{longtable}
\usepackage{multirow,booktabs}

\usepackage{supertabular}

\def\actaa{Acta Astron.}

\slugcomment{To appear in The Astrophysical Journal Letters (ApJL)}

\shorttitle{Stellar activity mimics a habitable-zone planet around Kapteyn's star}
\shortauthors{Robertson et al.}

\begin{document}


\title{Stellar activity mimics a habitable-zone planet around Kapteyn's star}


\author{Paul Robertson$^{1,2}$}

\author{Arpita Roy$^{1,2,3}$}
 
\author{Suvrath Mahadevan$^{1,2,3}$}

\altaffiltext{1}{Department of Astronomy and Astrophysics, The Pennsylvania State University}
\altaffiltext{2}{Center for Exoplanets \& Habitable Worlds, The Pennsylvania State University}
\altaffiltext{3}{The Penn State Astrobiology Research Center, The Pennsylvania State University}




\begin{abstract}
Kapteyn's star is an old M subdwarf believed to be a member of the
Galactic halo population of stars. A recent study has claimed the
existence of two super-Earth planets around the star based on radial
velocity (RV) observations. The innermost of these candidate
planets--Kapteyn b ($P = 48$ days)--resides within the circumstellar
habitable zone. Given recent progress in understanding the impact of
stellar activity in detecting planetary signals, we have analyzed the
observed HARPS data for signatures of stellar activity. We find that
while Kapteyn's star is photometrically very stable, a suite of spectral
activity indices reveals a large-amplitude rotation signal, and we
determine the stellar rotation period to be $143$ days. The spectral
activity tracers are strongly correlated with the purported RV signal of
``planet b,'' and the $48$-day period is an integer fraction ($1/3$) of
the stellar rotation period. We conclude that Kapteyn b is not a planet
in the Habitable Zone, but an artifact of stellar activity.

\end{abstract}

\section{\bf Introduction}

M dwarfs are very attractive targets in the search for potentially habitable terrestrial planets.  Their comparatively small masses and radii make the detection of low-mass planets in the habitable zone \citep[HZ;][]{kopparapu13} via radial velocity (RV) and transit photometry feasible even now.  The observational advantages of M dwarf planets also makes them amenable to atmospheric characterization \citep[e.g. GJ 1214b;][]{kreidberg14}, and there is a possibility that JWST could place constraints on the atmospheric composition--including biosignatures--of a terrestrial planet in the HZ of a nearby M dwarf \citep{seager13,cowan15}.

An insidious problem for RV detection of low mass planets around M dwarfs, however, is the presence of RV signals created by stellar activity.  Starspots, plage, filaments, convective suppression, and activity cycles can all create apparent RV shifts by altering the balance between red- and blueshifted portions of the stellar surface \citep[e.g.][and references therein]{dumusque14}.  Activity-induced RV signals appear preferentially at the stellar rotation period and its integer fractions \citep{boisse11}, so for old M stars with typical rotation periods around 100 days \citep{engle11}, periods associated with the HZ will be plagued by activity signals.  We have recently shown examples \citep{robertson14a,robertson14b} in which RV periodicities believed to be low-mass planets in or near the HZ were actually caused by activity.  Furthermore, for Barnard's star \citep{kurster03} and GJ 581 \citep{robertson14a}, the activity signals which drove RV shifts did not create a detectable photometric periodicity, suggesting that certain magnetic phenomena--probably magnetic suppression of convection--may only be apparent in spectral activity tracers.  Magnetic activity signals appear to be ubiquitous at the RV $\leq 1$ m/s level.

Kapteyn's star (= GJ 191) is a halo star with high proper motion, which may originate from outside the Galaxy \citep[see review by][]{kotoneva05}.  At 3.9 pc from the Sun, it is the nearest halo star.  The star is a target of the HARPS exoplanet search program, and \citet{bonfils13} reported preliminary solutions of possible planet detections.  More recently, \citet[][hereafter A14]{ae14} reported the detection of two super-Earth exoplanets at $P = 48.6$ (planet b) and $P = 121.5$ (planet c) days, based partially on those same HARPS observations.  The proposed orbit for planet b places it in the HZ.  Assuming a stellar age typical of halo stars, the existence of an HZ planet around Kapteyn's star allows for the fascinating possibility of life nearly as old as the Universe itself.  Considering the ramifications of such a discovery, it is doubly important to ensure Kapteyn b truly exists.

A14 considered the possibility that activity creates signals in the RVs of Kapteyn's star.  However, two of the activity tracers they examined--the Ca II H\&K lines ($S_{HK}$) and $V$-band photometry--are known to be suboptimal for M dwarfs \citep[e.g.][]{kurster03,diaz07,gds11}.  Here, we have reexamined activity on Kapteyn's star using the H$\alpha$ and Na I D lines.  We find that these lines offer a more significant detection of the $P \sim 140$-day periodicity observed by A14 in $S_{HK}$ and FWHM.  Based on the improved significance of the detection, and the mass and likely age for the star, we conclude the $143$-day periodicity must be the stellar rotation period.  This signal is strongly correlated with the proposed RV signal at $P=48$ days.  We conclude that the stellar rotation induces an RV signature which, when observed at the sparse cadence of the present RV set, creates the signal attributed to planet b.

\begin{figure*}
\begin{center}
\includegraphics[width=1.6\columnwidth]{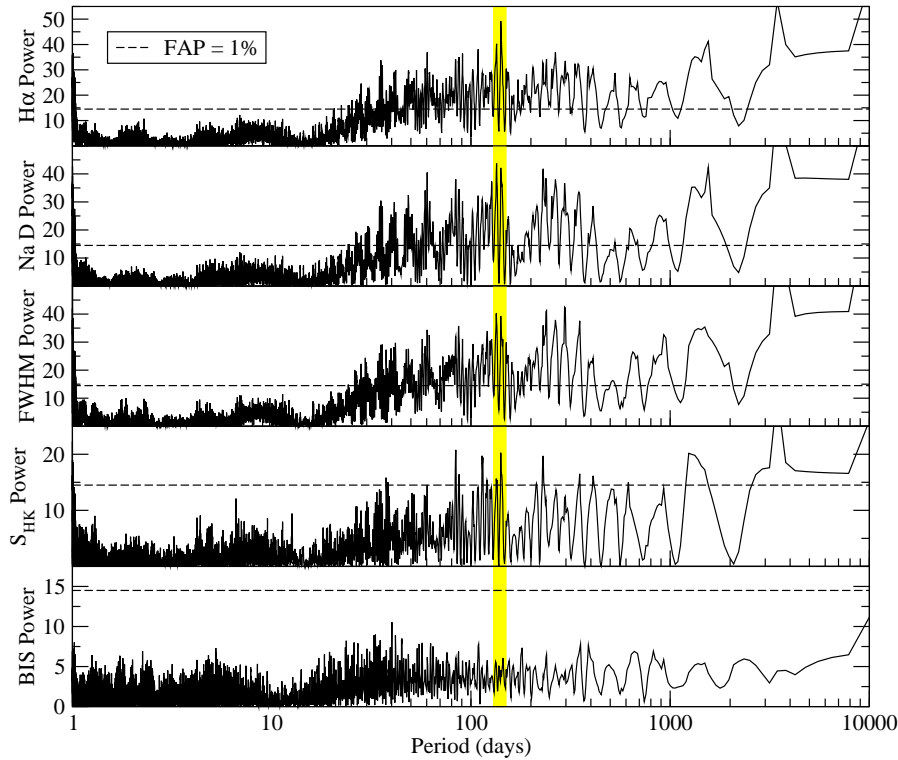}
\caption{\label{fig:act_ps}
\footnotesize Periodograms of the HARPS spectral activity tracers.  The vertical scales are not identical.  The peak near $P = 140$ days (highlighted) is present in every tracer except BIS.  The dashed lines indicate the power required for a FAP of 1\%.%
}
\end{center}
\end{figure*}

\section{\bf Data}

A14 claimed the detection of planets Kapteyn b and c based on RVs derived from 135 spectra obtained from the HARPS, HIRES, and PFS spectrographs.  Our analysis focused on the 95 HARPS spectra, which are all publicly available\footnote{\footnotesize Data obtained from the ESO Science Archive Facility under request numbers 115242 and 149297}.  We also considered the 20 (of 32) HIRES spectra where extracted spectra containing H$\alpha$ were available in the Keck Observatory Archive.  PFS spectra are not publicly available, and only account for 8 of the available RVs.  We adopted RVs from A14, which are derived using the HARPS-TERRA algorithm \citep{aebutler12} and a HIRES iodine-RV algorithm using these same spectra.

We considered an array of stellar activity indicators for Kapteyn's star.  For the HARPS spectra, we measured the activity indices for the Na I D resonance lines \citep[$I_{\textrm{D}}$;][]{gds11} and the H$\alpha$ line \citep[$I_{\textrm{H}\alpha}$;][]{robertson13}, which have been shown to be particularly sensitive to activity in M dwarf stars.  We also consider the $S_{HK}$ values provided in A14.  For HIRES spectra, only $I_{\textrm{H}\alpha}$ is available.

While the HARPS reduced data archive provides values for the line-shape characteristic quantities Bisector Inverse Slope (BIS) and full-width at half-maximum (FWHM), we have computed our own values for these activity tracers using the HARPS CCF (cross-correlation function) files, which are a product of the cross-correlation between stellar spectra and a tailored numerical mask. As in \citet{robertson14a}, we summed the individual CCFs of orders in the wavelength range between $\sim$5400 {\textendash} 6200 \AA~to obtain an ``average" CCF, then calculated BIS and FWHM according to \citet{Wright:2013}.  Although the periodicites and correlations observed herein for BIS and FWHM have higher significance when using our reduction, their qualitative behaviors are identical for the original HARPS values.

In addition to quantities derived from the HARPS spectra, we examined photometry of Kapteyn's star from the Hipparcos \citep{esa97} and ASAS \citep{pojmanski97} archives.  Photometry of Kapteyn's star from KELT-South \citep{pepper12} also exists, but the star's proper motion is so high that it moves between pixels throughout the observations, causing large offsets on the same timescale as the stellar rotation period (J. Rodriguez, private comm.).  We therefore do not incorporate the KELT-S data into this analysis.

We find that $S_{HK}$ and $I_{\textrm{D}}$ are clearly correlated with each other, and anticorrelated with H$\alpha$.  There are three spectra\footnote{\footnotesize BJD = 2454879.556, 2456424.446, 2456428.462} for which the spectral indices do not follow these correlations.  We excluded these spectra from our analysis, as they likely correspond either to flare events, other abnormal activity, or errors in the measurement of one or more lines.

\begin{figure*}
\begin{center}
\includegraphics[width=1.6\columnwidth]{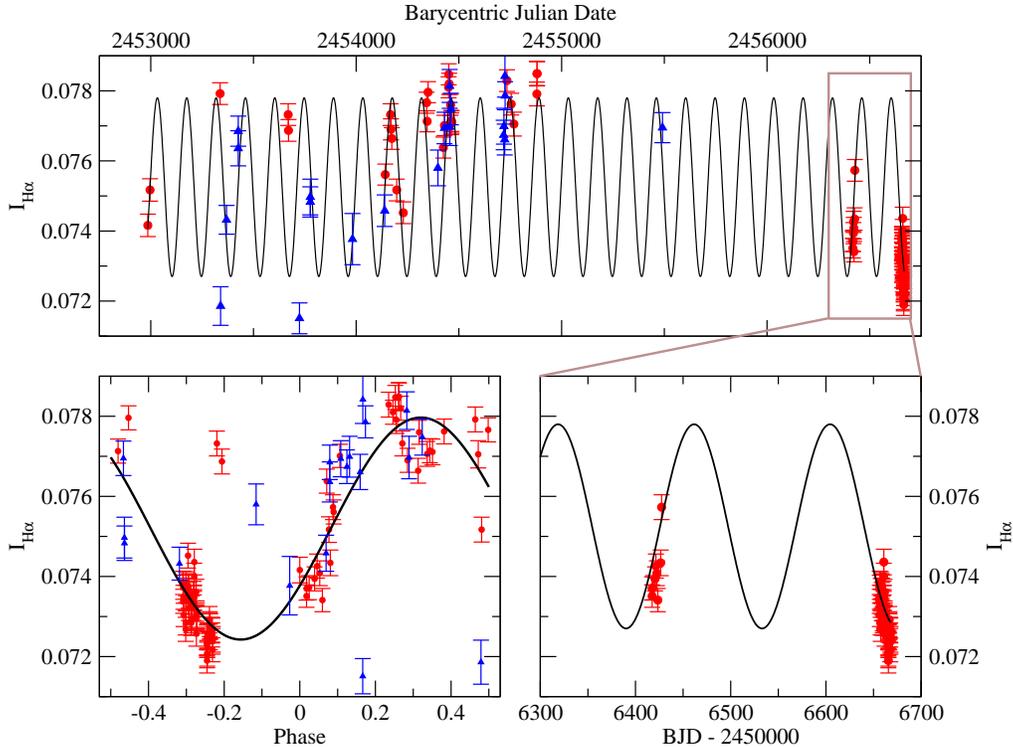}
\caption{\label{fig:rotation}
\footnotesize \emph{Top}: $I_{\textrm{H}\alpha}$ measurements from HARPS (red) and HIRES (blue), with our fit to the 143-day stellar rotation signal (black).  The inset (\emph{lower right}) shows a detailed view of the observations conducted by A14.  \emph{Lower left}: $I_{\textrm{H}\alpha}$ measurements folded to the rotation period.%
}
\end{center}
\end{figure*}

\section{\bf Analysis}

\subsection{Determining the stellar rotation period}

We began our analysis by examining the time series of the activity indices for any periodic signals which might leave an imprint on the RVs.  In particular, for the periods of the proposed planets, stellar rotation is the most likely source of astrophysical noise to create impostor signals.  We computed the generalized Lomb-Scargle periodogram \citep{zk09} to search for periodicities in the activity tracers.

In Figure \ref{fig:act_ps}, we show that the H$\alpha$, Na I D, FWHM, and $S_{HK}$ indices all show power near a period of 140 days, as observed by A14 for FWHM and $S_{HK}$.  The signal is strongest in $I_{\textrm{H}\alpha}$ (Figure \ref{fig:rotation}), where its periodogram power of $49.0$ corresponds to a false-alarm probability FAP $\sim 5 \times 10^{-18}$ according to the approximation given in Equation 25 of \citet{zk09}.  Periodograms of the activity indices also show power at very long periods, potentially indicative of a long-term trend or curvature; indeed, visual inspection of the activity series appears to show strong curvature as might be expected for a long-period cycle.  However, as explained further below, we believe this is likely a spurious artifact caused by sparse sampling of the stellar rotation signal.

We modeled sinusoids to the activity indices, using 140 days as an initial guess for the period.  For the combined HARPS/HIRES $I_{\textrm{H}\alpha}$ series, we found a best-fit period $142.9 \pm 0.3$ days with amplitude $0.0029 \pm 0.0002$ (= 4\%).  Our model to the data is shown in Figure \ref{fig:rotation}.  Fits to other tracers produced very similar results.

There are a number of outliers to this model, which may be caused by atmospheric distortion, errant data reduction, phase shifts in the rotation signal, or stochastic variability of the star.  Of particular note are the two HARPS observations from October 2005 (phase $\sim-0.22$); the anomalous behavior is observed twice in 2 nights, so the phenomenon is likely intrinsic to the star.  The anticorrelation between H$\alpha$ and the other tracers holds, precluding a flare.  While these excursions are interesting, and may merit further study, the overall behavior of the stellar activity is well described by the 143-day rotation signal.

\begin{figure*}
\begin{center}
\includegraphics[width=1.6\columnwidth]{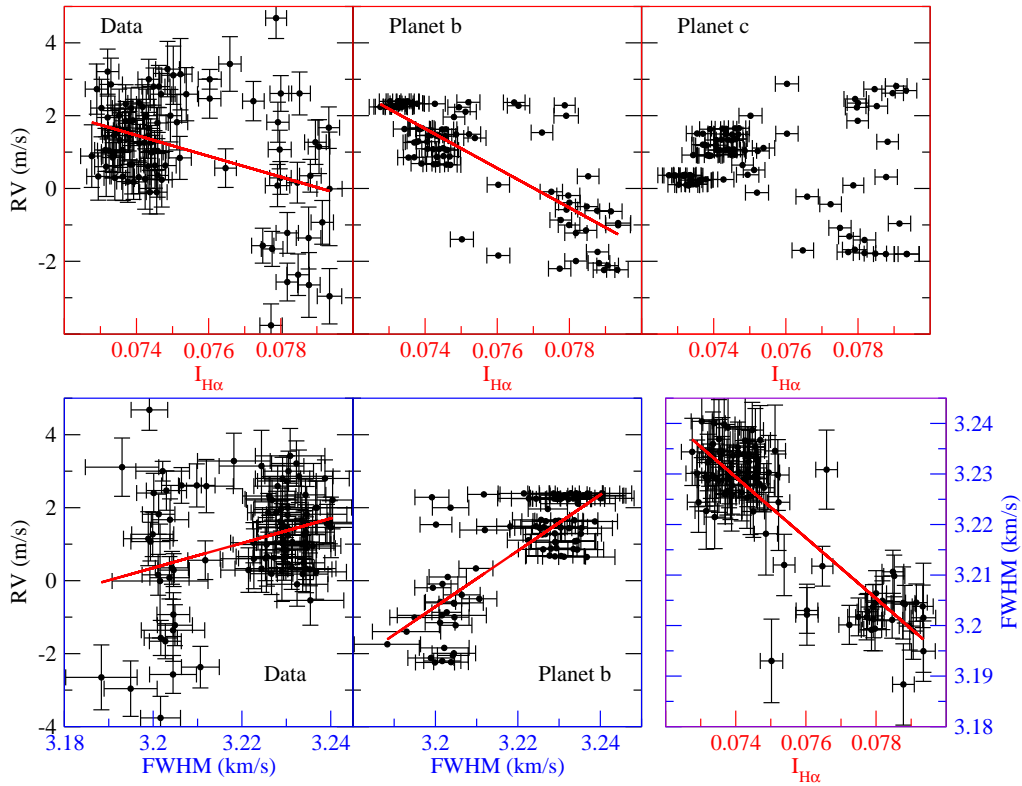}
\caption{\label{fig:act_rv}
\footnotesize RV as a function of activity for the HARPS-TERRA RVs and the predicted RVs from planets b and c.  We do not show uncertainties for the model-predicted RVs.  Our fits to correlated quantities are shown as red lines.  We also show FWHM as a function of $I_{\textrm{H}\alpha}$.  The strong anticorrelation confirms that stellar activity (traced by $I_{\textrm{H}\alpha}$) alters the stellar line profiles (traced by FWHM), thereby shifting RVs.%
}
\end{center}
\end{figure*}

It is necessary to justify our choice of a 143-day sinusoid over a polynomial trend for our model to the activity indices.  First, we note that the periodogram power at long periods is driven almost entirely by the two observing runs contributed by A14 in late 2013/early 2014, for which the activity indices are higher/lower (depending on the tracer) than the earlier values.  If--as suggested by our 143-day model (Figure \ref{fig:rotation}, inset)--those observations coincidentally occurred near the same extreme phase of a large-amplitude rotation signal, it could create the appearance of a long-term trend.  On a more physically motivated level, we note that at $M_* = 0.281 M_{\odot}$ \citep{segransan03}, Kapteyn's star is below the threshold \citep[$M_* = 0.35 M_{\odot}$;][]{chabrier97} where stars become fully convective.  Long-term surveys of activity in M stars \citep[e.g.][]{robertson13} have confirmed that fully convective stars, which therefore lack the $\alpha \Omega$ magnetic dynamo believed to drive Solar-type activity cycles, tend not to exhibit long-period magnetic cycles.  

Additionally, while the activity periodograms show many statistically significant peaks, we find that after fitting a 143-day sinusoid, the periodogram of the residuals shows no significant peaks, suggesting the power at other periods is simply aliasing from the rotation.  When modeling other periods, especially long-term trends, considerable power remains in the residuals.  We therefore conclude that a single 143-day sinusoid driven by the stellar rotation is the simplest, most physically plausible model for the observed variability.

\subsection{Impact of activity on RV}

A rotation period of 143 days for Kapteyn's star is a source of concern for both of the purported planets in the system.  The rotation period is uncomfortably close to the 121.5-day period of planet c, and almost exactly three times the period of planet b.  We therefore considered possible correlations between the RVs and stellar activity.

In Figure \ref{fig:act_rv} we show the HARPS RVs (which are the dominant source of information for the planet signals) plotted against $I_{\textrm{H}\alpha}$.  Performing a linear regression, we find a Pearson correlation coefficient $r = -0.38$.  The $r$-value suggests a probability $P(r) = 9 \times 10^{-5}$ that we would observe this level (or more) of correlation for 92 measurements of these quantities if $I_{\textrm{H}\alpha}$ and RV were in fact uncorrelated.

\citet{ae15} have questioned our use of residual RVs when considering activity-RV correlations.  While we disagree with their assertion that our statistical methods invalidate the results of our analysis of GJ 581 \citep[see][]{robertson15}, we seek to avoid any statistical ambiguity here.  Instead of modeling and subtracting planets or activity-RV correlations one at a time from the velocities and examining the residual velocities, we instead consider correlations between activity and the \emph{modeled} RV signatures of each planet as reported by A14.  If the predicted behavior of either planet matches the observed activity variations, it must be discarded as an activity signal, regardless of the statistical methods used to identify it.

Figure \ref{fig:act_rv} shows the predicted RV contribution of each planet at the time of each observation as a function of the corresponding $I_{\textrm{H}\alpha}$ measurement.  For planet c, its RVs are uncorrelated with activity.  On the other hand, the RVs of ``planet b" are very strongly anticorrelated with $I_{\textrm{H}\alpha}$.  Our fit to the RVs of this signal as a function of $I_{\textrm{H}\alpha}$ is qualitatively similar to what we obtained using the original velocities, but with less scatter, suggesting the activity-RV correlation is mostly contained within the model to the 48-day signal, and the 121-day signal distorts the relationship.  Fitting out stellar activity with a planetary orbit model at 48 days would explain the assertion in A14 that Kapteyn's star is an especially low-jitter target.

Other activity tracers show a qualitatively similar correlation with the predicted RVs of the 48-day signal.  The only difference is that $I_{\textrm{H}\alpha}$ and BIS evince a negative correlation, while the other indicators are positively correlated.  The fact that we observe simultaneous changes in both BIS and FWHM ensures that the line profiles are varying asymmetrically, and alter RVs.  It is especially reassuring to see (Figure \ref{fig:act_rv}) that FWHM is correlated with RV and the spectral line indices, confirming that activity is disorting the stellar line profiles--and, by extension, the RVs--rather than some external phenomenon which affects both quantities similarly.  In Table \ref{tab:correlations} we show the Pearson correlation coefficients of all spectral tracers with each planet signal as well as the original data.  The strong, consistent correlations offer conclusive evidence that the 48-day RV signal is not a planet, but a harmonic of the 143-day stellar rotation period.

\begin{table*}
\footnotesize
\centering

\begin{tabular}{l | c | c c c}
Tracer & Variable & Fit & $r$ & $P(r)$ \\
\hline & & & & \\
 & Data & $v_r = 23_{\pm 5} - 290_{\pm 70} \times I_{\textrm{H}\alpha}$ & $-0.38$ & $9 \times 10^{-5}$ \\
H$\alpha$ & Planet b (model) & $v_r = 42_{\pm 3} - 540_{\pm 40} \times I_{\textrm{H}\alpha}$ & $-0.79$ & $< 10^{-8}$ \\
 & Planet c (model) & $v_r = 7_{\pm 4} - 90_{\pm 60} \times I_{\textrm{H}\alpha}$ & $-0.16$ & $0.06$ \\
 & & & & \\
 & Data & $v_r = -5_{\pm 2} + 130_{\pm 40} \times I_{\textrm{D}}$ & $0.37$ & $1 \times 10^{-4}$ \\
Na I D & Planet b (model) & $v_r = -12_{\pm 1} + 260_{\pm 20} \times I_{\textrm{D}}$ & $0.80$ & $< 10^{-8}$ \\
  & Planet c (model) & $v_r = 0_{\pm 1} + 10_{\pm 30} \times I_{\textrm{D}}$ & $0.05$ & $0.32$ \\
  & & & & \\
  & Data & $v_r = -110_{\pm 40} + 30_{\pm 10} \times FWHM$ & $0.31$ & $0.001$ \\
FWHM (km s$^{-1}$) & Planet b (model) & $v_r = -250_{\pm 20} + 77_{\pm 7} \times FWHM$ & $0.77$ & $< 10^{-8}$ \\
  & Planet c (model) & $v_r = -30_{\pm 30} + 9_{\pm 9} \times FWHM$ & $0.11$ & $0.15$ \\
  & & & & \\
  & Data & $v_r = -4_{\pm 2} + 18_{\pm 6} \times S_{HK}$ & $0.32$ & $9 \times 10^{-4}$ \\
Ca II H\&K & Planet b (model) & $v_r = -6_{\pm 1} + 27_{\pm 5} \times S_{HK}$ & $0.52$ & $5 \times 10^{-8}$ \\
  & Planet c (model) & $v_r = 0_{\pm 1} + 2_{\pm 5} \times S_{HK}$ & $0.04$ & $0.35$ \\
  & & & & \\
  & Data & $v_r = 0_{\pm 1} - 0.05_{\pm 0.04} \times BIS$ & $-0.16$ & $0.06$ \\
BIS (m s$^{-1}$) & Planet b (model) & $v_r = 0.6_{\pm 0.5} - 0.10_{\pm 0.03} \times BIS$ & $-0.31$ & $0.001$ \\
  & Planet c (model) & $v_r = 0.3_{\pm 0.4} - 0.0_{\pm 0.9} \times BIS$ & $-0.10$ & $0.17$ \\
\hline
\end{tabular}

\caption{\label{tab:correlations}
\footnotesize Linear fits, Pearson correlation coefficients ($r$), and $P$-values for HARPS RVs (in m s$^{-1}$) as a function of the spectral activity tracers.  Here, ``Data" refers to the original RVs, while the planet models are the predicted RVs at each observation for each planet according to A14.}
\end{table*}

\subsection{Unresponsive activity tracers}

We find it remarkable that the stellar rotation signal and its correlation with RV appear so consistently across four different activity tracers.  However, it is equally important to point out that we see neither periodic variability nor activity-RV correlations in the photometry from Hipparcos and ASAS, and do not recover the periodicity in BIS.  The absence of features in these data are consistent with the findings of A14.  A power spectrum of the ASAS photometry shows the peaks near 1 year and 1100 days mentioned in A14, but we find that those peaks match peaks from the window function (the periodogram of the sampling), confirming they are caused by poor sampling.

The lack of a rotation signal in the photometry is not surprising, as previous studies of M dwarfs \citep{kurster03,robertson14a} have revealed magnetic phenomena in H$\alpha$ which alter RVs without a corresponding photometric signature.  Bisectors are more difficult to measure in M type stars due to the prodigious presence of line blends, and the lower HARPS signal-to-noise compared to FGK stars \citep{Basturk:2011}, which likely explains both the lack of significant periodicity and the less significant BIS-RV correlations.

\section{\bf Discussion}

What caused the appearance of a planet-like RV signal at 48 days?  The sampling cadence of the HARPS time series for Kapteyn's star is insidious, allowing power from the rotation signal to leak to shorter periods.  The unfortunate coincidence that the two 2013/2014 observing runs both sampled the rotation signal near the same phase worsened matters, as it reduced the significance of the 140-day signal in both RV and the activity tracers.  The presence of another RV signal at 121 days further obscures the velocity contribution from the stellar rotation.  

\begin{figure}
\begin{center}
\includegraphics[width=0.98\columnwidth]{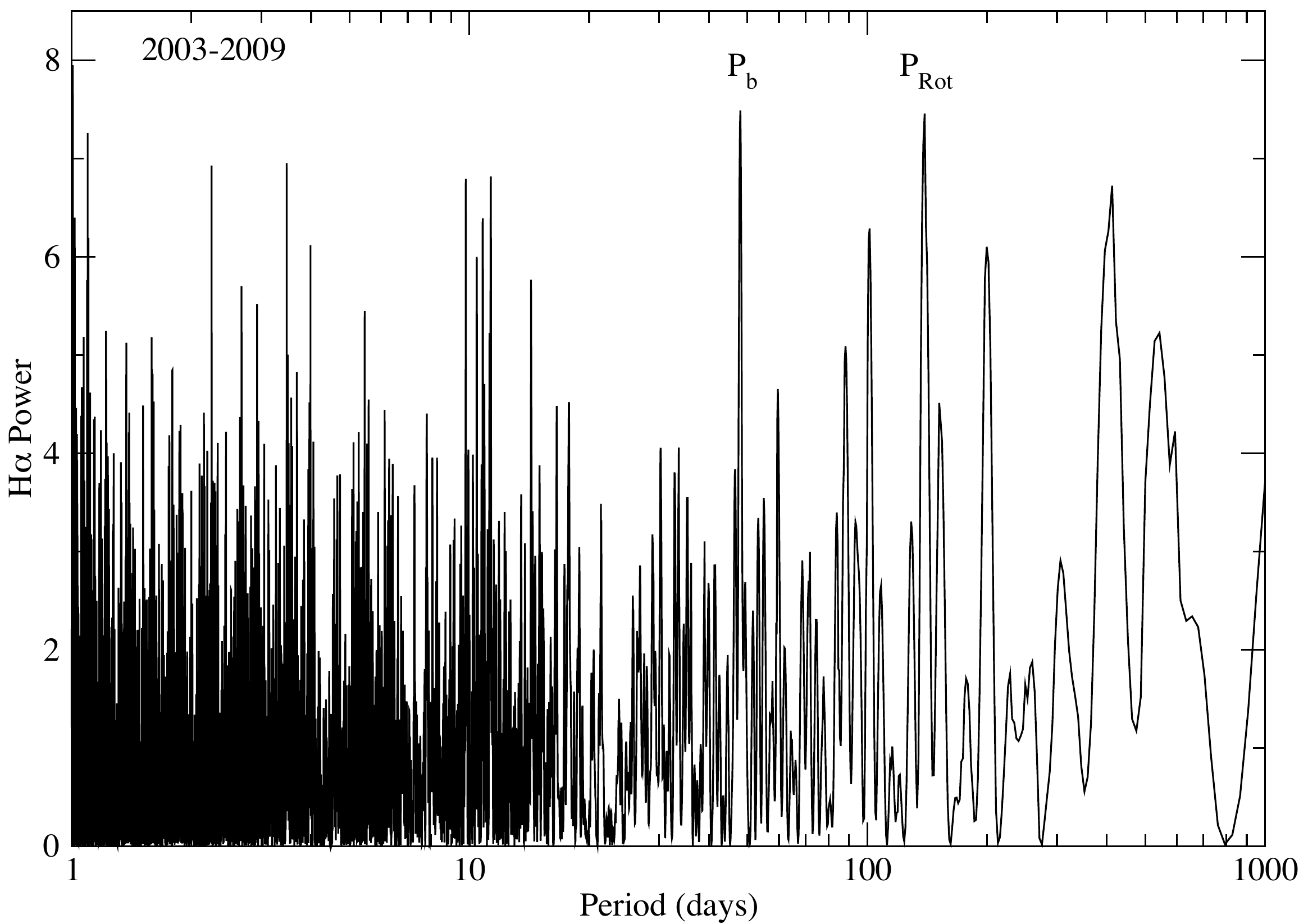}
\caption{\label{fig:ha_early}
\footnotesize Periodogram of $I_{\textrm{H}\alpha}$ for HARPS observations from 2003-2009.  Peaks corresponding to the stellar rotation period ($P_{rot}$ = 143 days) and its harmonic at $P_{rot}/3$--the period of ``planet b" ($P_b = 48$ days)--are clearly distinguished.%
}
\end{center}
\end{figure}

\citet{bonfils13} identified a 2-signal RV solution very similar to that of A14 using only the HARPS observations from 2003-2009.  Even for the complete data, most of the phase coverage of the 48-day signal comes from these observations; the new data from A14 only cover approximately a third of the 48-day phase curve.  While the 2003-2009 data alone do not offer statistically significant detections of any periodicities, they are sufficient to see that the rotation signal and its harmonic at $P_{rot}/3$--i.e.~the period of ``planet b"--are present during this epoch.  In Figure \ref{fig:ha_early}, we show the periodogram of $I_{\textrm{H}\alpha}$ from 2003-2009.  The $P_{rot}/3$ harmonic is the strongest peak in the power spectrum.  In other words, for the first 30 HARPS observations, which are crucial to the appearance of the 48-day signal, stellar activity matches the periodicity of the ``planet."  The deleterious effect of sampling in this data set illustrates an important requirement for ultra-precise RV surveys searching for low-mass planets; a consistent, regular cadence is required to resolve, model, and correct semi-periodic stellar signals in RV data.

As to the origin of the 121-day period, we see little in our analysis to contradict the interpretation of that signal as a planet.  Planets are known to exist around low-mass halo stars, as evidenced by the multi-planet system around Kepler-444 \citep{campante15}, so it is reasonable that planets would exist around Kapteyn's star.  However, its period is worryingly close to the stellar rotation period and its aliases, and is equally vulnerable to the sampling concerns outlined for the 48-day signal.  Kapteyn c should probably be considered a planet candidate pending additional observations to confirm the physical origin of the 121-day signal.

We direct particular attention to the stark contrast between the strengths of the activity-RV correlations in $I_{\textrm{H}\alpha}$ and $I_{\textrm{D}}$ to those in the other spectral tracers.  This work, alongside numerous previous studies \citep[e.g.][]{kurster03,gds11,robertson13,robertson14a} have shown that activity tracers redder than $S_{HK}$--particularly H$\alpha$--often reveal activity signals in old M stars that do not appear in Ca H\&K.  Too frequently, these unrecognized activity signals have been interpreted instead as low-mass planets in the HZ.  Enough examples have now been provided to conclude that an analysis of $S_{HK}$ alone is insufficient for ruling out false-positive planets around M dwarfs.  It is our opinion that either highly confident knowledge of the stellar rotation period, or an analysis of $I_{\textrm{H}\alpha}$, $I_{\textrm{D}}$, and/or an equivalently suitable M dwarf activity tracer should be required for the confirmation of low-mass planets around M stars.

Our measured 143-day rotation period for Kapteyn's star is consistent with its previous descriptions as an old halo star.  According to the age-rotation-$L_X$ curve from \citet{engle11}, this period suggests an age of approximately 10 Gyr.  We find the amplitude of the stellar rotation signal in the spectral tracers surprising for such an old, quiet star; with an amplitude of 4\% in $I_{\textrm{H}\alpha}$, it is more than twice the amplitude ($1.7$\%) of the rotation signal observed for GJ 581 \citep{robertson14a}.  Additional observations might determine whether the amplitude varies, or is a perpetually strong signal.  Also interesting is the fact that the extant data are consistent with the rotation signal staying in phase over the $\sim 10$-year span of the observations--more than 25 rotations--although again, the sampling makes it impossible to be sure this is the case.

Kapteyn's star joins Barnard's star \citep{kurster03} and GJ 581 \citep{robertson14a} on a growing list of old M stars for which one or more activity signals are observed in H$\alpha$ (and potentially other spectral tracers) and create RV shifts, but have not been detected in existing photometry.  \citet{kurster03} attributed this behavior to localized regions on the stellar surface where strong magnetic fields inhibit convection, but do not create dark spots.  This phenomenon deserves additional attention, as it now appears to be common in M dwarfs, and clearly poses a significant obstacle to RV detection of exoplanets in those systems.

\section{\bf Conclusions}

We have analyzed the stellar activity of Kapteyn's star using a variety of activity indicators.  We found the star has a rotation period of approximately 143 days, and creates a periodic signal in the $I_{\textrm{H}\alpha}$, $I_{\textrm{D}}$, FWHM, and $S_{HK}$ indices.  This signal is present in the RVs, and due to the sampling of the extant time series, creates a planet-like signal at 48 days ($= P_{rot}/3$).  The results of this analysis, along with recent work on GJ 581 and GJ 667C highlight a number of precautions which must be taken to avoid false positives when searching for low-mass M dwarf planets with RV.  For Kapteyn's star, the signal of the claimed planet ``c" does not seem to be correlated with stellar activity indicators, and we recommend continued observations to confirm the planetary nature of this signal.

\begin{acknowledgements}
We thank J. Rodriguez, T. Beatty, R. Siverd, and J. Pepper for access to KELT-South photometry.  We are also grateful to the anonymous referee for a prompt, thorough, and insightful report.  We acknowledge support from NSF grants AST-1006676, AST-1126413, and AST-1310885; the
Penn State Astrobiology Research Center; the NASA Astrobiology Institute (NNA09DA76A); and the Center for Exoplanets and Habitable Worlds (CEHW). The CEHW is supported by the Pennsylvania State University, the Eberly College of Science, and the Pennsylvania Space Grant Consortium.  This research makes use of the Keck Observatory Archive, operated by the Keck Observatory and NExScI, under contract with NASA. 
\end{acknowledgements}

\end{document}